# Identifying and reducing interfacial losses to enhance color-pure electroluminescence in blue-emitting perovskite nanoplatelet light-emitting diodes

Robert L. Z. Hoye,*,†,# May-Ling Lai,‡,# Miguel Anaya,‡ Yu Tong,‖,§ Krzysztof Gałkowski,‡,‖ Tiarnan Doherty,‡ Weiwei Li,† Tahmida N. Huq,† Sebastian Mackowski,‖ Lakshminarayana Polavarapu,‖,§ Jochen Feldmann,‖,§ Judith L. MacManus-Driscoll,† Richard H. Friend,‡ Alexander S. Urban,¥,§ Samuel D. Stranks*,‡

† Department of Materials Science and Metallurgy, University of Cambridge, 27 Charles Babbage Rd, Cambridge CB3 0FS, UK

‡ Cavendish Laboratory, University of Cambridge, JJ Thomson Ave, Cambridge CB3 0HE, UK

‖ Chair for Photonics and Optoelectronics, Nano-Institute Munich, Department of Physics, Ludwig-Maximilians-Universität München, Königinstraße 10, 80539 Munich, Germany

§ Nanosystems Initiative Munich (NIM) and Center for NanoScience (CeNS), Schellingstraße 4, 80799 Munich, Germany

¥ Nanospectroscopy Group, Nano-Institute Munich, Department of Physics, Ludwig-Maximilians-Universität München, Königinstraße 10, 80539 Munich, Germany

‼ Institute of Physics, Faculty of Physics, Astronomy and Informatics, Nicolaus Copernicus University, 5th Grudziadzka St., 87–100 Toruń, Poland

AUTHOR INFORMATION

**Corresponding Author**

*Robert L. Z. Hoye, Email: rlzh2@cam.ac.uk

*Samuel D. Stranks, Email: sds65@cam.ac.uk

**Author Contributions**

# These authors contributed equally

**Abstract**

Perovskite nanoplatelets (NPls) hold great promise for light-emitting applications, having achieved high photoluminescence quantum efficiencies (PLQEs) approaching unity in the blue wavelength range, where other metal-halide perovskites have typically been ineffective. However, the external quantum efficiency (EQE) of blue-emitting NPl light-emitting diodes (LEDs) have only reached 0.12%, with typical values well below 0.1%. In this work, we show that the performance of NPl LEDs is primarily hindered by a poor electronic interface between the emitter and hole-injector. Through Kelvin Probe and X-ray photoemission spectroscopy measurements, we reveal that the NPls have remarkably deep ionization potentials (≥6.5 eV), leading to large barriers for hole injection, as well as substantial non-radiative decay at the interface between the emitter and hole-injector. We find that an effective way to reduce these non-radiative losses is by using poly(triarylamine) interlayers. This results in an increase in the EQE of our blue LEDs emitting at 464 nm wavelength to 0.3%. We find that our results can be generalized to thicker sky-blue-emitting NPls, where we increase the EQE to 0.55% using the poly(triarylamine) interlayer. Our work also identifies the key challenges for further efficiency increases.

**TOC GRAPHICS**

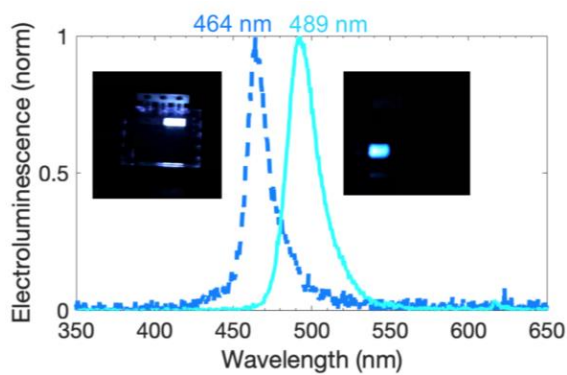

MAIN TEXT

Lead-halide perovskites have recently garnered significant interest, owing to rapidly rising efficiencies in photovoltaics and light emitting diodes (LEDs).[1] These performances are mainly due to high internal luminescence quantum yields,[2–4] and long diffusion lengths.[5] The exceptional optoelectronic properties of perovskites can be realized in polycrystalline thin films with higher defect densities than Si or GaAs, suggesting a tolerance to defects.[6–8] This allows high-performing films to be obtained through a variety of low-temperature routes, such as solution-processing or thermal evaporation.[2,9] In particular, the band gap of lead-halide perovskites can be simply tuned over the entire visible light spectrum, from 1.55 eV to 3.2 eV, by changing the halide from iodide to bromide to chloride.[10] Although this is appealing for light-emitting applications, a critical limitation is that the photoluminescence quantum efficiencies (PLQEs) drop off strongly in the blue wavelength region. Thus, while near-infrared (iodide) and green-emitting (bromide) perovskites have achieved PLQEs close to unity,[11,12] the PLQEs of archetypical blue-emitting $CsPbBr_xCl_{3-x}$ nanocrystals are typically ≤10%.[13,14] While the best iodide-based and bromide-based perovskite LEDs (PeLEDs) have reached external quantum efficiencies (EQEs) above 20%,[12,15] PeLEDs emitting at wavelengths less than 475 nm exhibit EQEs at least an order of magnitude lower.[13,16] Blue-emitters are critical for applications in solid-state white lighting and displays.[13] For displays in particular, primary blue emitters with wavelengths <475 nm are required to meet National Television System Committee (NTSC) standards,[13] while color-pure, saturated emission at 465 nm is required to meet the International Telecommunications Union Radiocommunication Sector (ITU-R) standards for ultrahigh definition displays.[17]

Several efforts have been made to increase the PLQE of blue-emitting perovskites. In inorganic lead bromide-chloride nanocrystals emitting at 470 nm, Congreve *et al.* investigated Mn doping. Adding 0.19% Mn increased the PLQE from 9% to 28%, which translated to improvements in EQE from 0.5% to 2.1%.[13] This process critically required fine control over the Mn content, since higher Mn content resulted in a roll-off in PLQE due to energy transfer to the Mn-ions, as well as emission at 600 nm wavelength.[13] Higher PLQEs have been achieved without the need for dopants using bromide-based two-dimensional perovskite nanoplatelets (NPls), which quantum confine excitons in one dimension.[18,19] It has been shown that fine atomic-level control over the NPl thickness can be achieved through facile solution synthesis, allowing the emission energy of $CsPbBr_3$ to be increased from 2.48 eV (nanocubes) to 3.2 eV by reducing the number of monolayers of corner-sharing $[PbBr_6]^{4-}$ octahedra present. By repairing surface vacancy defects present in these NPls, we have shown that a PLQE of (60 ± 4) % at 464 nm can be reached.[18] Wu *et al.* also recently achieved a PLQE of 96% at the same wavelength by passivating halide vacancies in similar cesium lead bromide NPls.[16] However, early efforts to make devices from perovskite NPls resulted in EQEs <0.006%, with broad emission spanning from 425 nm to 520 nm wavelength.[19] Wu *et al.* recently improved the color-purity and the EQEs up to 0.12%,[16] but more typical values for NPl PeLED EQEs are well below 0.1%.[18,20] Thus, although perovskite NPls have exhibited some of the highest PLQEs amongst blue perovskite emitters, these have not been reflected in similarly-impressive device performances. A key parameter that has not been addressed in previous works on blue perovskite NPls is the influence of interfaces with charge injector materials. In particular, these interfaces can act as sites of efficiency losses due to non-radiative recombination or inefficient charge injection,[1] and this problem is compounded by the lack of clarity on the band structure of blue-emitting perovskite NPls.[21]

In this work, we systematically identify the role of each interface in our NPl LEDs on non-radiative losses through photoluminescence (PL) quantum efficiency measurements. We rationalize these losses by determining the NPl band structure with macroscopic Kelvin Probe and X-ray photoemission spectroscopy. We introduce a poly(triarylamine) interlayer to our PeLEDs and find that this significantly improves performance. Through the analysis of the band structure, single-carrier devices and time-resolved PL, we rigorously probe the role played by the poly(triarylamine) interlayer. The NPls we focus on are those emitting at 464 nm wavelength (blue), but also make comparison to NPls emitting at 489 nm (sky-blue) to generalize our findings among blue perovskite NPls. We identify the key limiting interface, highlighting that further improvements will need careful management of the hole injecting interface to allow efficient injection into deep energy levels, as well as minimizing non-radiative losses and charge quenching.

We synthesized CsPbBr$_3$ NPls using our previously-reported reprecipitation method[18] (detailed in the Supporting Information) and spin-cast these colloidal solutions onto glass substrates. The absorbance of the NPl thin films exhibited excitonic peaks at wavelengths of 454 nm (blue NPls) and 467 nm (sky-blue NPls), with the PL peaks exhibiting Stokes shifts to 460 nm (blue) and 487 nm (sky-blue) wavelengths, respectively (Figure 1a, Table 1). The size of these Stokes shifts are comparable to our previous measurements on colloidal NPls,[18] as well as the Stokes shifts measured in perovskite nanocrystals.[22,23] By fitting the absorbance measurements with Elliot's model, we obtained an absorption onset of 2.87 eV (blue) and 2.72 eV (sky-blue) from our NPl thin films (Figure S1, Supporting Information). We performed transmission electron microscopy (TEM) measurements on NPls deposited onto copper TEM grids. The NPls imaged top-down were asymmetric (Figure 1b), indicating that they were *a*-axis oriented. We measured the NPl thicknesses to be (1.7 ± 0.1) nm (blue) and (3.1 ± 0.2) nm (sky-blue), which are consistent with 3 and 5 monolayers of [PbBr$_6$]$^{4-}$ octahedra in the NPls respectively, with each monolayer being 0.59 nm thick.[18] The emission wavelengths, full width at half maximum of the PL peaks and absorption onsets we obtained in the thin films here are comparable with the number of monolayers we measured in our previous work focusing on NPls in colloidal solution.[18] We also deposited our NPls onto copper grids covered with PEDOT:PSS (PD; Figure S2, Supporting Information) and the NPls remained *a*-axis oriented.

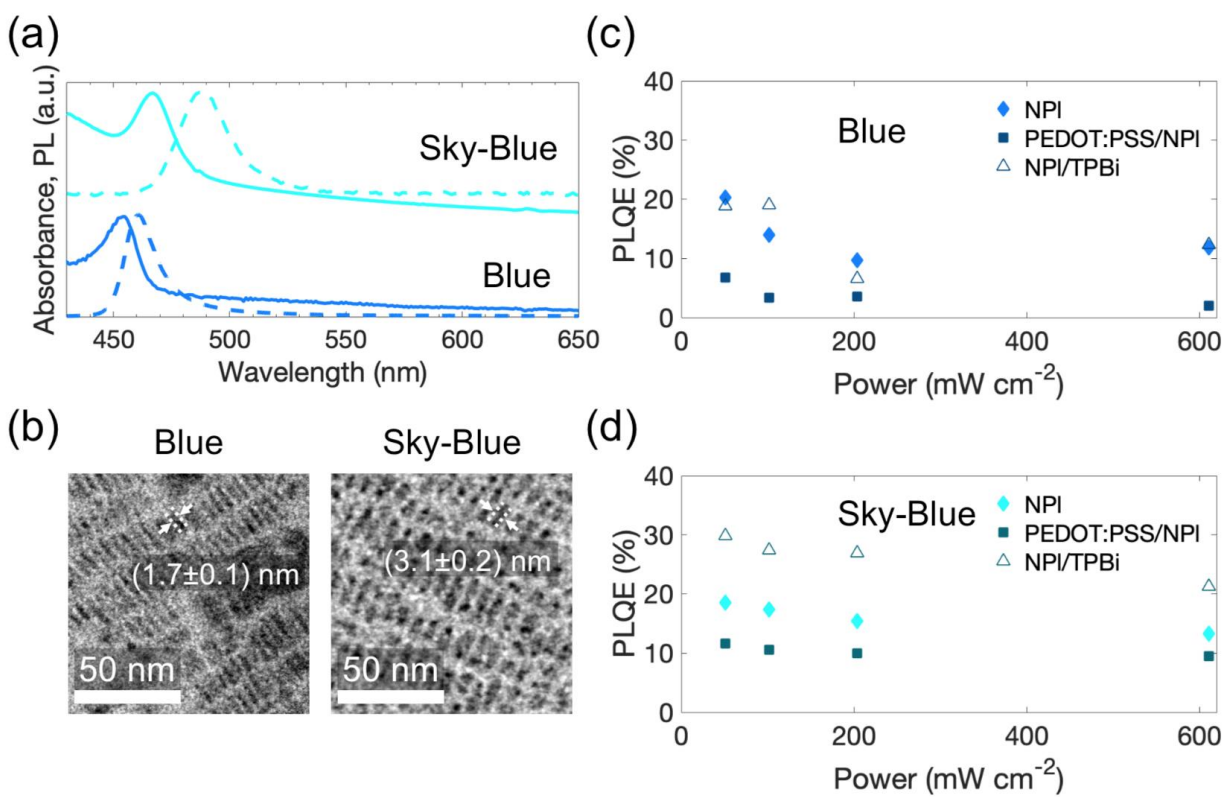

**Figure 1**. Optical properties and structure of blue and sky-blue CsPbBr$_3$ perovskite nanoplatelets (NPls). (a) Absorbance (solid line) and photoluminescence (dashed line) measurements. (b) Transmission electron microscopy images of blue and sky-blue NPls. Photoluminescence quantum efficiency (PLQE) measurements as a function of excitation power with a continuous wave 405 nm laser of (c) blue and (d) sky-blue NPls deposited on glass or PEDOT:PSS, compared to NPls deposited on glass and with 35 nm TPBi evaporated on top. Data points are the average of three samples.

We investigated the impact of conventional LED charge-injection layers on the PLQE of the NPl emitters. For the hole-injector, we used PD, and for the electron-injector, we used 2,2',2"-(1,3,5-Benzinetriyl)-tris(1-phenyl-1-H-benzimidazole) (TPBi). The PLQEs of the NPl samples were measured inside an integrating sphere using the method described by de Mello, *et al.*[24] The NPls were excited by a 405 nm wavelength continuous laser, for which a power density of 100 mW cm$^-$ gives a similar flux density of carriers as at maximum luminance in our best performing devices (approximately $1.4 \times 10^{17}$ electron s$^{-1}$ cm$^{-2}$, which corresponds to 20 mA cm$^{-2}$; see later in Figure 3b). Under 100 mW cm$^{-2}$, the neat blue NPl thin films exhibited a PLQE of 14% (Figure 1c), which is comparable to that of our previous colloidal solutions.[18] However, when the blue NPls were spin-cast onto PD layers, we found that the PLQE was reduced to 3%. This quenching of the PLQE was consistent over the measured range of excitation power densities (Figure 1c), indicating the presence of significant non-radiative recombination at the interface between the NPls and the PD hole-injector. By contrast, a layer of TPBi evaporated over the blue NPl thin film on glass had a negligible effect on the PLQE (Figure 1c). The sky-blue NPls showed a similar trend for neat films and for those deposited on PD, where the emission yields were also significantly quenched by the PD layer (Figure 1d; Table 1). However, in this case, the layer of TPBi over the sky-blue NPls strongly increased the PLQE from 17% to 27% at 100 mW cm$^{-2}$ (Figure 1d, Table 1).

**Table 1**. Effect of contacts on the PLQE of the NPl emitters when excited with a 405 nm wavelength continuous laser with a power density of 100 mW cm$^{-2}$

| NPl | PL peak wavelength (nm) | PL FWHM (nm) | PLQE (%) | | |
|---|---|---|---|---|---|
| | | | On glass | On PD | With TPBi |
| Blue | 460 | 16 | 14 | 3 | 19 |
| Sky-blue | 487 | 25 | 17 | 11 | 27 |

To rationalize the effects of PD and TPBi on the PLQE of the NPl emitters, particularly the increase in PLQE we observed with TPBi, we determined the band positions of the NPls using X-ray photoemission spectroscopy (XPS) and macroscopic Kelvin Probe measurements. The Kelvin Probe measurements allowed us to obtain stabilized work function measurements and avoid the surface charging effects we observed when we attempted ultraviolet photoemission spectroscopy measurements (data not shown). The valence band spectra of the NPls measured by XPS are shown in Figure 2a,b. The binding energy scale was calibrated to the C 1s peak for adventitious carbon (284.8 eV) and the valence band spectrum was acquired immediately afterwards. By monitoring the Pb 4f levels immediately before and 3 h after the measurement of the valence band spectra, we found there to be only marginal changes in binding energy, peak intensity or peak shape (Figure S3, Supporting Information), consequently showing there to be negligible charging effects or beam damage under these conditions. To determine the valence band ($E_{VB}$) to Fermi level ($E_F$) offset ($E_{VB} - E_F$), we fitted the density of states of $CsPbBr_3$ (obtained from Ref. 25), convoluted with a Gaussian representing instrument broadening, to the leading edge of the valence band spectra.[8,26] From this, we found $E_{VB} - E_F$ to be 1.9 eV for the blue NPls and 1.5 eV for the sky-blue NPls. We used Kelvin Probe measurements to determine the work function (details in the Experimental Section of the Supporting Information), and found the values to be 4.9 eV for blue NPls and 5.0 eV for sky-blue NPls. From these measurements, the ionization potentials of the NPls are 6.8 eV (blue) and 6.5 eV (sky-blue). Using the absorption onsets obtained from Elliot modelling of the absorption spectra (described earlier and in Figure S1, Supporting Information), we found the electron afffinities of the NPls to be 3.9 eV (blue) and 3.8 eV (sky-blue). We note that the measured band positions for these materials are deep. Although the $E_{VB} - E_F$ values are similar to those previously-reported for bulk $CsPbBr_3$ thin films, the work function is higher.[25] This may be due to changes in the work function (or indeed $E_{VB} - E_F$ as well) as the band gap is increased from bulk $CsPbBr_3$ to quantum-confined NPls, or it could be due to the effects of the ligands (*e.g.*, surface dipoles).[27] However, since both XPS and Kelvin Probe are surface-sensitive techniques, we expect the band positions we measured to reflect those that influence charge injection at the interfaces of the NPls with the contacts.

We constructed a band diagram of the NPls sandwiched between the hole and electron injectors in Figure 2c using the energy levels of the PD and TPBi from literature.[19,28] The work function of the electrodes connected to these are also shown (4.8 eV for indium tin oxide, or ITO, for holes;[8,19] 3 eV for calcium for electrons[29]). It can be seen from Figure 2c that there is a large (>1.3 eV) offset between the hole injection level and ionization potential of the emitter. In spite of this potential barrier to hole injection, there is precedent for hole injection still occurring based on wide band gap CdSe quantum dot LED systems, in which holes can be injected across a barrier of almost 2 eV due to an Auger-assisted process across a type II heterojunction.[30] However, excitons generated through charge-injection or photoexcitation can be easily dissociated at the interface between PD and NPls, and recombine non-radiatively. This will be further exaggerated if this interface contains additional defects. By contrast, the lower LUMO of TPBi to the electron affinity of the NPls confines photogenerated excitons within the NPl layer from this side. It is also possible that the TPBi may passivate the top surface of the NPls, giving rise to the higher PLQEs observed in the sky-blue NPls (*cf.* Figure 1d). We note that while there is a hole-injection barrier between the sky-blue NPls and TPBi, the HOMO of TPBi is approximately level with the valence band maximum of the blue NPls and holes may not be as effectively confined within the emitter. This may explain why the PLQE was not enhanced in the blue NPls with a TPBi overlayer (*cf.* Figure 1c).

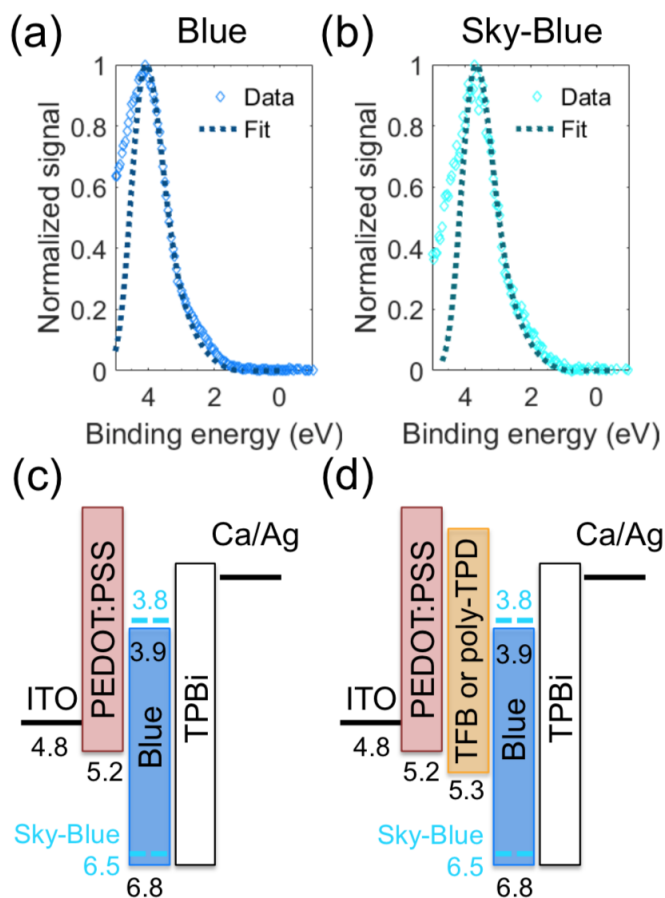

**Figure 2**. X-ray photoemission spectroscopy (XPS) measurements of the valence band spectra of (a) blue NPls on TFB and (b) sky-blue NPls on poly-TPD on ITO-coated glass. Dashed lines are fits from the calculated density of states. Band structure of perovskite LED devices with blue or sky-blue emitters (c) without or (d) with a polymer interlayer. The band positions of ITO, PD, TPBi and Ca were obtained from literature.[8,19,28,29,31,32]

Our PLQE, XPS and Kelvin Probe results highlight that the problematic quenching region is the PD/NPl interface. When we tested our NPls sandwiched between PD and TPBi contacts in full LEDs, electroluminescence was weak and the low EQEs (Figure 3a–c; Table 2) were on a similar order of magnitude to previous reports using PD as the hole-injector.[19] To address this, we investigated the effect of the addition of a wide band gap polymer interlayer between the NPls and PD (Figure 3d). We investigated two poly(triarylamines): poly(9,9-dioctylfluorene-alt-*N*-(4-sec-butylphenyl)diphenyl)-diphenylamine (TFB) and poly(*N*,*N*'-bis(4-butylphenyl)-*N*,*N*'-bisphenylbenzidine) (poly-TPD). We found that the highest color-purity and performance was obtained using TFB hole-injecting interlayers for the blue emitters, and poly-TPD hole-injecting interlayers for the sky-blue emitters (Figure S4, S6 and Table S1, Supporting Information). By using these poly(triarylamine) interlayers, we increased the EQE by almost two orders of magnitude for both the blue NPls (from 0.007% to 0.3%) and sky-blue NPls (from 0.004% to 0.24%), along with increases in luminance and luminous efficiency (Figure 3b,c, Table 2). The orders of magnitude improvements exhibited by our champion devices were also found in the mean performances of 5–8 devices measured for each condition (Table S2, Supporting Information).

We would expect further increases in performance by improving the NPls, such as through passivation. To test this, we introduced $PbBr_2$ complexed with oleylamine and oleic acid ($PbBr_2$-ligand) to the colloidal solution of our sky-blue NPls. Our previous work showed that the addition of $PbBr_2$-ligand passivates surface traps on the colloidal NPls, leading to an enhancement in the PLQE, along with longer PL lifetimes, but without any change to the spectral shape.[18] By passivating our NPl films, we here find that the EQE of our champion sky-blue LEDs doubled to 0.55% (Figure 3c, Table 2; Figure S7, Supporting Information), which is consistent with a further reduction in non-radiative recombination in the emitter. The mean EQE of 8 sky-blue NPl PeLEDs doubled when using passivated sky-blue NPl emitters instead of unpassivated ones (Table S2, Supporting Information), with the increase in EQE exceeding the uncertainty, showing the improvement in performance to be statistically significant. We note that our champion performance is over five times larger than the EQE previously reported for $CsPbBr_3$ NPl PeLEDs emitting at 480 nm wavelength that did not have passivated NPl emitters.[20]

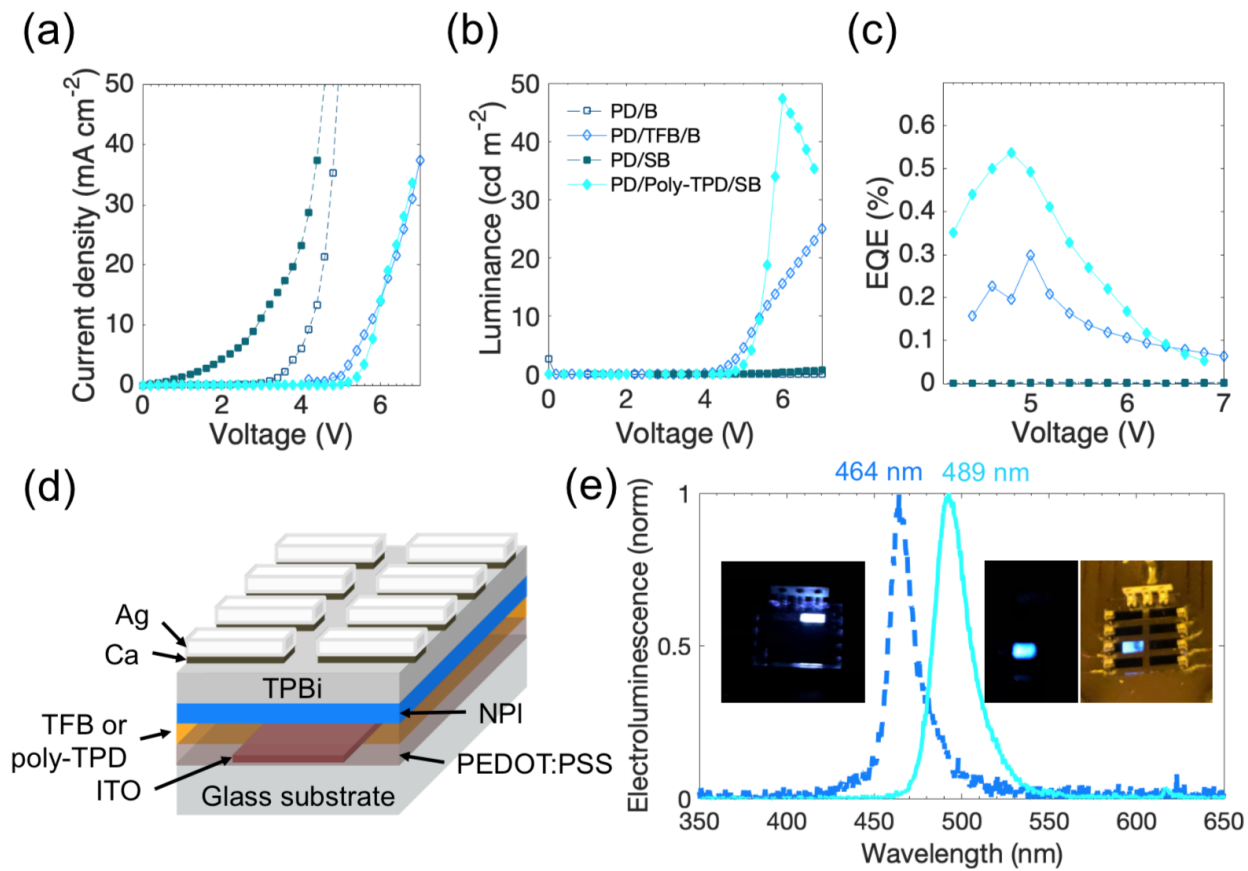

**Figure 3**. Performance of NPls in LED device structures. (a) Current density, (b) luminance and (c) external quantum efficiency (EQE) of champion devices on PEDOT:PSS (PD) or on PD with a polymer interlayer. B = blue-emitting NPl, SB = sky-blue-emitting NPl. (d) Device structure. (e) Electroluminescence spectra with photographs of the devices under operation inset.

**Table 2**. Performance metrics of champion perovskite NPl devices. The reproducibility of the devices are shown in Table S2 in the Supporting Information. The number of decimal points the values are quoted to here are consistent with the uncertainties obtained from measuring multiple samples (see Table S2).

| NPl | Hole injector | EL $\lambda$ (nm) | EL FWHM (nm) | Peak radiance (W sr$^{-1}$ m$^{-2}$) | Peak luminous efficiency (cd A$^{-1}$) | Peak luminance (cd m$^{-2}$) | Peak EQE (%) |
|---|---|---|---|---|---|---|---|
| Blue | PD | - | - | 0.20 | 0.007 | 25 | 0.007 |
| Blue | PD/TFB | 464 | 16 | 0.3 | 0.3 | 40 | 0.3 |
| Sky-Blue | PD | - | - | 0.01 | 0.007 | 3 | 0.004 |
| Sky-Blue | PD/poly-TPD | 489 | 26 | 0.48 | 0.48 | 120 | 0.24 |
| Sky-Blue (+PbBr$_2$) | PD/poly-TPD | 487 | 21 | 0.13 | 1.1 | 32 | 0.55 |

N.B.: $\lambda$ = wavelength, FWHM = full width at half maximum

The electroluminescence spectra of our devices are given in Figure 3e, with photographs of the devices under operation inset. It can be seen that the wavelengths of the electroluminescence peaks from our devices matched the wavelengths of the PL peaks from the non-passivated films to within a few nanometers (Table 1 *cf.* Table 2). The color-purity of the emission from our LEDs was improved over the earliest report on electroluminescence from perovskite NPls[19] because we were able to achieve films with uniform nanoplatelet thickness that only emitted at one wavelength by finely controlling the molar ratio of the precursors (Cs$_2$CO$_3$/PbBr$_2$) and volume of the antisolvent. The full width at half maximum (FWHM) of the electroluminescence spectrum of the blue NPl devices was 16 nm, and 21 nm for the sky-blue NPl devices (Table 2). These are in close agreement with the FWHM of our PL measurements (Table 1) and sharper than standard organic emitters (*e.g.*, TFB, >80 nm FWHM), inorganic LEDs based on GaN (30 nm FWHM),[33] and also blue quantum dot emitters (30 nm FWHM),[31] and are amongst the narrowest reported for blue PeLEDs.[16] In addition, the electroluminescence wavelength (464 nm) and FWHM of our blue NPl PeLEDs match the requirements set by ITU-R for ultrahigh definition displays (466 nm wavelength; 20 nm FWHM).[17] We also checked the stability of the emission from our devices over the operating range. Each electroluminescence spectrum was taken within 30 s of applying the bias. Our blue-emitting PeLEDs were color-pure over the operating rang, maintaining their low FWHMs, with only a small increase in TFB emission at wavelengths below 450 nm for higher biases (Figure S11a, SI). This would not have significantly impacted the EQE or luminance because the shoulder in TFB emission was small and the photopic factor at that wavelength is very low. Our sky-blue PeLEDs remained color-pure over the entire measured voltage range of 6 V to 8 V, and the FWHMs of the electroluminescence peaks remained unchanged (Figure S11b, SI). Since the emission from the poly(triarylamine) layer in the PeLEDs is largest at the highest voltages, we measured the electroluminescence spectrum of our PeLEDs with passivated sky-blue NPls at 8 V (Figure S11c, SI). This was color-pure apart from a small shoulder of emission from poly-TPD at wavelengths below 460 nm, which, again, would have a very small influence on EQE or luminance. We also note that we previously showed our perovskite NPls to be stable in air and under UV illumination, with the PL peak maintaining the same wavelength and FWHM.[18] However, the electroluminescence intensity of our blue and sky-blue PeLEDs drops to half of the original intensity within 1 min of continuous bias at 5 V, which is consistent with the other current state-of-the-art for blue PeLEDs, and future efforts to improve device stability would need to be made.[21]

We next performed detailed measurements to understand why the poly(triarylamine) interlayers led to the orders of magnitude improvement in EQEs that we obtained. Two possibilities are: (1)

a reduction in non-radiative recombination or (2) an increase in charge-injection efficiency. Both polymers emit broadly over the blue wavelength range, and their PL spectra overlap with those of the NPls, thus preventing us from measuring the change in PLQE with the use of these interlayers. To investigate the effect of the polymer interlayer on charge-injection efficiency, we made hole-only (ITO/PD/(polymer)/NPl/MoO$_x$/Au) and electron-only (ITO/Al:ZnO/PEI/NPl/TPBi/Ca/Ag) devices, and measured the current injection from the PD/(polymer) and TPBi. We found that in both blue and sky-blue NPl devices, the electron current density was higher than the hole current density (Figure 4a,b). This is consistent with better electron injection from TPBi to the NPl (than hole injection at the other interface) owing to a negligible barrier to electron injection at the TPBi/NPl interface. The current densities of the bipolar devices in Figure 3 are overall lower than the electron current densities and, based on our single-carrier devices, are primarily limited by hole-injection. For the sky-blue LEDs in particular, the bipolar current density is very similar to the hole current density. In both blue and sky-blue NPls, the addition of a polymer interlayer reduced the bipolar and hole current densities (Figure 3a,4a,b). If the effect of the polymer layer on device performance were primarily due to a reduced hole-injection barrier, we would have expected an increase in the hole current density, which was not the case. In addition, the turn-on voltage for light emission was unchanged, remaining at approximately 4 V (Figure 3b). These results are consistent with the analysis of the band structure, in which the reduction in the hole-injection barrier by 0.1-0.2 eV through the use of the polymer interlayers is small compared to the size of the barrier (Figure 2c,d).

To understand the influence of the poly(triarylamine) layer on interface recombination, we performed time-correlated single photon counting (TCSPC) measurements of the decay in the PL of the NPl layer with different interfaces. We used a monochromator to measure emission at a single wavelength (464 nm for blue emitters; 485 nm for sky-blue emitters). Due to the polymer emission in a similar spectral region, we decoupled the PL decay of the NPl layer from that of the polymer through a fitting procedure utilizing measurements of the PD/polymer and PD/polymer/NPl stacks (details in Figure S8, Supporting Information). As depicted in Figure 4c, we found that the blue NPls on PD had a faster initial PL decay than NPls on glass. This is consistent with the quenching of the PLQE by PD (*cf.* Figure 1). However, the extracted PL decay of the NPls on PD/TFB had a slower decay than on PD alone and was similar to the decay of the NPls on glass. Very similar trends in PL decay were also obtained from the sky-blue emitters (Figure 4d). Our time-resolved PL results therefore indicate that the polymer interlayer reduced interface recombination. A reduction in non-radiative recombination due to the polymer interlayer is consistent with the reduction in hole current density in our LEDs (Figure 4a,b), since this is the sum of carriers recombining in the emitter and non-radiatively recombining at the interface. In addition, this is also consistent with a previous study on perovskite photovoltaics that revealed significant surface recombination at the interface with PD (even though device leakage was absent), which can be reduced by replacing the PD with poly-TPD.[34]

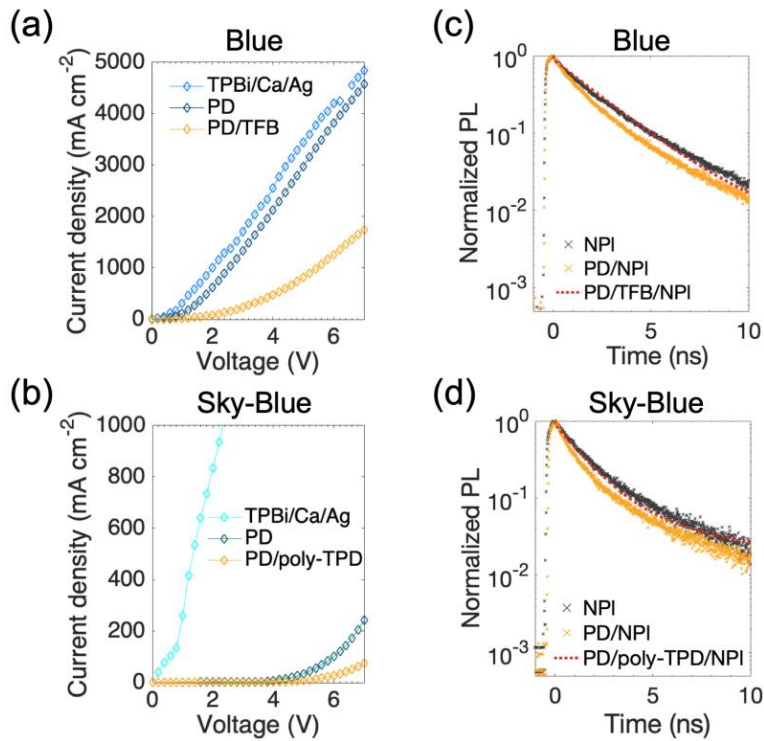

**Figure 4**. Single-carrier device data for (a) blue-emitting and (b) sky-blue-emitting NPls. The current density from electron-only devices injecting from the TPBi/Ca/Ag layer is compared to hole-only devices injecting from PEDOT:PSS (PD) or PD with a polymer interlayer. Close-up photoluminescence (PL) decays of (c) blue-emitting and (d) sky-blue-emitting NPls drop-cast onto different substrates, measured at emission wavelengths of 464 nm and 485 nm, respectively. The full PL decays are shown in Figure S8, SI. The PL decay of the NPl on polymer was extracted through fitting (detailed in Figure S8, SI), owing to the emission of the polymer at the same wavelengths. The PL decay measurements were made using time-correlated single photon counting with a 407 nm wavelength excitation laser with a repetition rate of 20 MHz and fluence of 7.4 nJ cm$^{-2}$ pulse$^{-1}$ (blue NPls) or 9 nJ cm$^{-2}$ pulse$^{-1}$ (sky-blue NPls). Fluence-dependent measurements are shown in Figure S9,10, SI. We note that the PL decays in parts c & d were normalized to the peak PL value for each sample.

Importantly, our single-carrier and band position measurements suggest that the key limiting factors that must now be overcome to further improve the EQEs of blue-emitting NPl PeLEDs are: (i) reducing the charge imbalance and (ii) reducing the hole-injection barrier. Both issues arise in part due to the deep band positions of our NPls (Figure 2c,d), resulting in hole injection being more difficult than electron injection, which would have also resulted in the electron current density exceeding the hole current density. These issues could be solved by changing the ligands to effect shallower band positions through ligand-induced surface dipoles,[27] or through judicious choice of the interface modifiers and hole-injectors with higher HOMO/ionization potential levels. Nevertheless, matching the high ionization potentials of the current NPls will be challenging and care needs to be taken that the modifiers/injectors do not lead to increased non-radiative recombination at the interface with the NPl emitter.

In conclusion, we found that one of the key process limiting the efficiencies of blue perovskite NPl devices is quenching of the PLQE at the interface with the PD hole injector. We found that this loss mechanism could be overcome by using a poly(triarylamine) interlayer (TFB for blue NPls; poly-TPD for sky-blue NPls) between the PD and emitter. In doing so, we increased the EQEs by almost two orders of magnitude for both the blue LEDs (to 0.3%) and sky-blue LEDs (to 0.55%). These EQE values are higher than previously reported for blue perovskite NPls, which only reached up to 0.12% despite their high PLQEs. Our analysis of single-carrier devices and time-resolved PL showed that the role of the poly(triarylamine) interlayers was primarily to reduce non-radiative recombination at the hole-injector interface. Our work pushes NPls forward as a viable contender for efficient blue PeLEDs, particularly since we were able to demonstrate color-pure emission at 464 nm (blue) and 489 nm (sky-blue) wavelength, with sharper room-temperature electroluminescence than conventional organic, inorganic and colloidal Cd-based quantum dot emitters. This will allow the important advantages of perovskite NPls to be ultimately exploited in display and lighting applications, namely the ability to achieve high PLQEs without the need for dopants, and an emission wavelength that can be finely tuned through facile solution-based methods. Our work also highlights that a key area that should be addressed to push efficiencies further is the effects of the deep band positions of the perovskite NPls, which give rise to large hole-injection barriers but negligible electron injection barriers.

ASSOCIATED CONTENT

**Supporting Information**. Experimental section, absorbance spectra fitting, X-ray photoemission spectroscopy measurements, supporting device measurements, supporting TCSPC measurements.

AUTHOR INFORMATION

**Corresponding Authors**

*Robert L. Z. Hoye, Email: rlzh2@cam.ac.uk

*Samuel D. Stranks, Email: sds65@cam.ac.uk

**Author Contributions**

# These authors contributed equally

S.D.S., R.L.Z.H. and A.S.U. conceived the project. S.D.S. and R.L.Z.H. supervised the project. M.-L.L. synthesized the blue NPls and devices. R.L.Z.H. synthesized the sky-blue NPls and performed the optical, photoluminescence and time-resolved photoluminescence characterization on all NPls, and also characterized and analyzed the single-carrier devices. R.L.Z.H. and M.A. made and characterized the sky-blue NPl LEDs. Y.T. and L.P. developed the recipe for synthesizing the blue and sky-blue NPls, and also helped to develop the device structure. K.G. performed the Kelvin Probe measurements. Y.T., L.P. and T.D. performed the TEM measurements. W.L. performed the XPS characterization. T.N.H. performed the AFM measurements. All authors contributed to writing the manuscript.

**Notes**

The authors declare no competing financial interest.

ACKNOWLEDGMENT

The authors thank Prof. Dr. David Egger and Dr. Javad Shamsi for useful discussions. R.L.Z.H. acknowledges support from the Royal Academy of Engineering under the Research Fellowship scheme (No.: RF\201718\17101), as well as support from Magdalene College, Cambridge. This project has received funding from the European Research Council (ERC) under the European Union's Horizon 2020 research and innovation programme (grant agreements No. 756962 [HYPERION] and No. 759744 [PINNACLE]). S.D.S acknowledges support from the Royal Society and Tata Group (UF150033). M.-L.L. and R.H.F. acknowledge support from EPSRC grant EP/M005143/1. Y.T., L.P., J.F. and A.S.U. acknowledge financial support from the Bavarian State Ministry of Science, Research, and Arts through the grant "Solar Technologies go Hybrid" (SolTech), and from the DFG through the grant "e-conversion" within the framework of the German Excellence Initiative. K.G. acknowledges support from the Polish Ministry of Science and Higher Education within the Mobilnosc Plus program (grant no. 1603/MOB/V/2017/0). T.D. acknowledges the National University of Ireland (NUI) for a Travelling Studentship. W.L. and J.L.M.-D. acknowledge support from EPSRC grant EP/L011700/1 and EP/N004272/1, as well as the Isaac Newton Trust (Minute 13.38(k)). T.N.H. acknowledges support from the EPSRC Centre for Doctoral Training in Graphene Technology (EP/L016087/1).REFERENCES

(1) Samuel D. Stranks; Robert L. Z. Hoye; Dawei Di; Richard H. Friend; Felix Deschler. The Physics of Light Emission in Halide Perovskite Devices. *Adv. Mater.* **2018**, *Early View*. https://doi.org/10.1002/adma.201803336.

(2) Brenes, R.; Guo, D.; Osherov, A.; Noel, N. K.; Eames, C.; Hutter, E. M.; Pathak, S. K.; Niroui, F.; Friend, R. H.; Islam, M. S.; et al. Metal Halide Perovskite Polycrystalline Films Exhibiting Properties of Single Crystals. *Joule* **2017**, *1* (1), 155–167. https://doi.org/10.1016/j.joule.2017.08.006.

(3) Richter, J. M.; Abdi-Jalebi, M.; Sadhanala, A.; Tabachnyk, M.; Rivett, J. P. H.; Pazos-Outón, L. M.; Gödel, K. C.; Price, M.; Deschler, F.; Friend, R. H. Enhancing Photoluminescence Yields in Lead Halide Perovskites by Photon Recycling and Light Out-Coupling. *Nat. Commun.* **2016**, *7*, 13941. https://doi.org/10.1038/ncomms13941.

(4) Abdi-Jalebi, M.; Andaji-Garmaroudi, Z.; Cacovich, S.; Stavrakas, C.; Philippe, B.; Richter, J. M.; Alsari, M.; Booker, E. P.; Hutter, E. M.; Pearson, A. J.; et al. Maximizing and Stabilizing Luminescence from Halide Perovskites with Potassium Passivation. *Nature* **2018**, *555*, 497–501. https://doi.org/10.1038/nature25989.

(5) Stranks, S. D.; Eperon, G. E.; Grancini, G.; Menelaou, C.; Alcocer, M. J. P.; Leijtens, T.; Herz, L. M.; Petrozza, A.; Snaith, H. J. Electron-Hole Diffusion Lengths Exceeding 1 Micrometer in an Organometal Trihalide Perovskite Absorber. *Science* **2013**, *342*, 341–345. https://doi.org/10.1126/science.1243982.22